\documentclass[showpacs,preprintnumbers,amsmath,amssymb,twocolumn]{revtex4}

  \usepackage[dvips]{graphics}

\usepackage{graphicx}

\begin{document}

\title{Optically-Induced Suppression of Spin Relaxation in Two-Dimensional
Electron Systems with Rashba Interaction}

\author{ Yuriy V. Pershin}

\affiliation{\small{Department of Physics and Astronomy, Michigan
State University, East Lansing, Michigan 48824-2320, USA}}

\begin{abstract}
A pulsed technique for electrons in 2D systems, in some ways
analogous to spin echo in nuclear magnetic resonance, is
discussed. We show that a sequence of optical below-band gap
pulses can be used to suppress the electron spin relaxation due to
the D'yakonov-Perel' spin relaxation mechanism. The spin
relaxation time is calculated for several pulse sequences within a
Monte Carlo simulation scheme. The maximum of spin relaxation time
as a function of magnitude/width of the pulses corresponds to
$\pi$-pulse. It is important that even relatively distant pulses
efficiently suppress spin relaxation.
\end{abstract}


\maketitle

There has been a lot of experimental \cite{r1,r2,r3,r4,r5} and
theoretical \cite{r7,r8,r9,r11,r12,r13,r14,r15,r151,r16} interest
in the physics of spin relaxation in semiconductor structures. The
main reason for that is the potential of spintronic applications
\cite{r17,r19,r20,r21,r22,r23,r24,r27}. Controlling spin
relaxation rate is interesting both from a fundamental and
practical point of view. One of the ways through which spin
polarization can be lost is spin-orbit interaction. Of particular
interest is Rashba spin-orbit (SO) interaction \cite{r31}, which
is observed in asymmetric heterostructures. Corresponding spin
relaxation mechanism is known as D'yakonov-Perel' (DP) spin
relaxation mechanism \cite{r32}.

Let us consider a system of 2D electrons confined in a quantum
well or heterostructure. The Rashba spin-orbit interaction can be
regarded as an effective momentum-dependent magnetic field acting
on electron spins. In the presence of the effective magnetic
field, the electron spins feel a torque and precess in the plane
perpendicular to the magnetic field direction with an angular
frequency $\overrightarrow{\Omega}( \overrightarrow{k} )$. This
precession leads to an average spin relaxation (dephasing).
Momentum scatterings reorient the direction of the precession
axis, making the orientation of the effective magnetic field
random and trajectory-dependent. Therefore, frequent scattering
events suppress the precession and consequently the spin
relaxation. This is the motional-narrowing behavior, accordingly
to which the spin relaxation time
$\tau_s^{-1}\propto\tau_p$\cite{r32}, where $\tau_p$ is a momentum
scattering time.

Spin echo is a standard way to overcome dephasing in nuclear
magnetic resonance experiments \cite{Abragham}. Nuclear spin
magnetization, after a free induction decay, can be restored, as a
result of the effective reversal of the dephasing of the spins
(refocusing) by the application of a refocusing RF pulse (applied
in a time shorter than or of the order of $T_2$ time).
Unfortunately, this method can not be directly applied to electron
spin coherence in heterostructures. One of the obstacles is that
the minimum achievable RF pulse length of $\sim10$ns is of the
order or even longer than the typical spin coherence time.
Moreover, the effective magnetic field due to SO interaction is
fixed only between two consecutive scattering events. Therefore, a
refocusing pulse sequence should have a pulse separation of the
order of $\tau_p$ and pulse duration much shorter than $\tau_p$.
In what follows we discuss a possible realization of such
refocusing pulse sequence based on a method from femtosecond
optics.

In this Letter we consider dynamics of electron spin polarization
in a two-dimensional semiconductor structure like a quantum well
or heterostructure under a train of intense optical below-band gap
circularly-polarized pulses. Recent experiments have been
demonstrated that an effective magnetic field due to an optical
below-band gap pulse coherently rotates electron spins on a time
scale of $\sim 150$fs \cite{awschalom}, which is much shorter than
typical values of $\tau_p$ in clean structures. The mechanism of
spin rotation is based on the optical Stark effect \cite{Ostark}.
Physically, the optical Stark effect in semiconductors is related
to optically-induced modification (dressing) of quantum states
\cite{Ostark}, including optically-induced spin splitting
\cite{awschalom,flatte}. Since below-band gap laser does not
excite real excitons, the optically-induced spin splitting lasts
only as long as the pump pulse. The purpose of the current
investigation is to study the effect of the pulse sequence on
electron spin relaxation time in 2D quantum structures with
dominant D'yakonov-Perel' spin relaxation mechanism. Electron spin
rotations due to the pulse sequence result in partial
compensations of spin precessions due to the Rashba interaction.
Correspondingly, electron spin relaxation time becomes longer.

\begin{figure}[bt]
\centering
\includegraphics[height=8cm,angle=270]{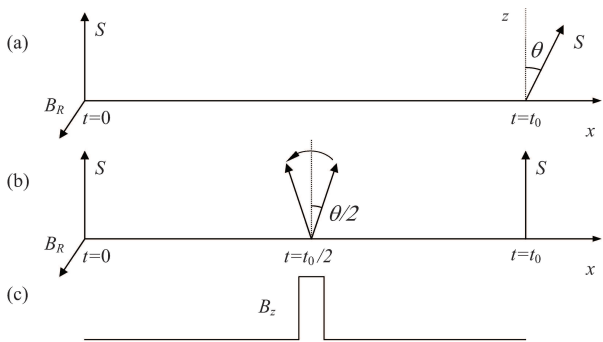}
\caption{The effect of an ideal  $\pi$-pulse on the electron spin
polarization vector $S$: (a) evolution of the spin polarization
vector without a pulse ; (b) evolution of the spin polarization
vector with a pulse ; (c) pulse profile.} \label{fig1}
\end{figure}

The main idea of our approach is illustrated in Fig. \ref{fig1}.
Let us consider the evolution of an electron spin (initially
aligned with $z$-axis) during a time interval between two
consecutive scattering events. Using a semiclassical approach to
electron space motion (the electrons are treated as classical
particles in the effective-mass approximation), we assume that an
electron moves along a straight trajectory with a constant
velocity. Fig. \ref{fig1}(a) shows that without a pulse, the
direction of electron spin at $t=t_0$ is changed by an angle
$\theta$ due to precession around the effective spin-orbit
magnetic field $B_R$. Fig. \ref{fig1}(b) demonstrates the effect
of the light pulse applied at $t=t_0/2$ (Fig. \ref{fig1}(c)) with
such a width and intensity that the electron spin rotates around
$z$-axis by angle $\pi$. It is readily seen that in this case at
$t=t_0$ the electron spin is directed in the initial
$z$-direction, so the effect of Rashba spin-orbit interaction is
eliminated. In reality, of course, it is not possible to apply
pulses exactly in the middle of each free flight interval for each
electron, hence, a residual relaxation remains.

In order to get a quantitative estimation of the effect, we
perform a Monte Carlo simulation of spin dynamics in the presence
of optical below-band gap pulses. The electron spin relaxation
time is calculated as a function of the electron spin precession
angle $\varphi$ (due to a pulse) for different selected values of
the spacing between pulses $T_B$ and for two types of pulse
sequences: unidirectional and alternating. For the sake of
simplicity, we assume that the effective magnetic field due to the
pulse is much stronger than the effective magnetic field due to
the spin-orbit interaction. This assumption allows us to consider
the electron spin precession events due to the pulses as
instantaneous.

Within a Monte Carlo simulation scheme, it is assumed that the
electrons move along trajectories, which are defined by bulk
scattering events (scattering on phonons, impurities, etc.), with
an average velocity $v$. The angular frequency corresponding to
the Rashba coupling can be expressed as
$\overrightarrow{\Omega}=\eta \overrightarrow{v} \times \hat{z}$,
where $\eta=2\alpha m^* \hbar^{-2}$, $m^*$ is the effective
electron mass, and $\alpha$ is the interaction constant that
enters into the Rashba spin-orbit coupling Hamiltonian
\begin{equation}
H_R=\alpha \hbar^{-1} \left( \sigma_x p_y-\sigma_y p_x \right).
\end{equation}
Here, $\overrightarrow{\sigma}$ is the Pauli-matrix vector
corresponding to the electron spin. The spin of a particle moving
ballistically over a distance $1/\eta$ will rotate by the angle
$\gamma=1$. The angle of the spin rotation per mean free path
$L_p$ is given by $\eta L_p $. It is assumed that at the initial
moment of time the electron spin are polarized in $z$-direction
(perpendicular to the plane) by a pump beam. We calculate $\langle
\overrightarrow{S} \rangle$ as a function of time by averaging
over an ensemble of electrons and taking into account both
Rahsba-induced and optically-induced spin precessions. The spin
relaxation time is evaluated by fitting the time-dependence of
$\langle \overrightarrow{S} \rangle$ to an exponential decay. The
detailed description of the basic Monte Carlo simulation scheme
can be found in Ref. \cite{r7}. We note that the selected Monte
Carlo algorithm correctly describes the physics of DP relaxation.
However, since all scattering parameters and temperature effects
are taken into account only via two parameters $L_p$ and $\tau_p$,
the temperature dependence as well as the role of Coulomb
scattering can not be easily evaluated, and more sophisticated
simulations \cite{r15} are required.

\begin{figure}[tb]
\centering
\includegraphics[height=8cm,angle=270]{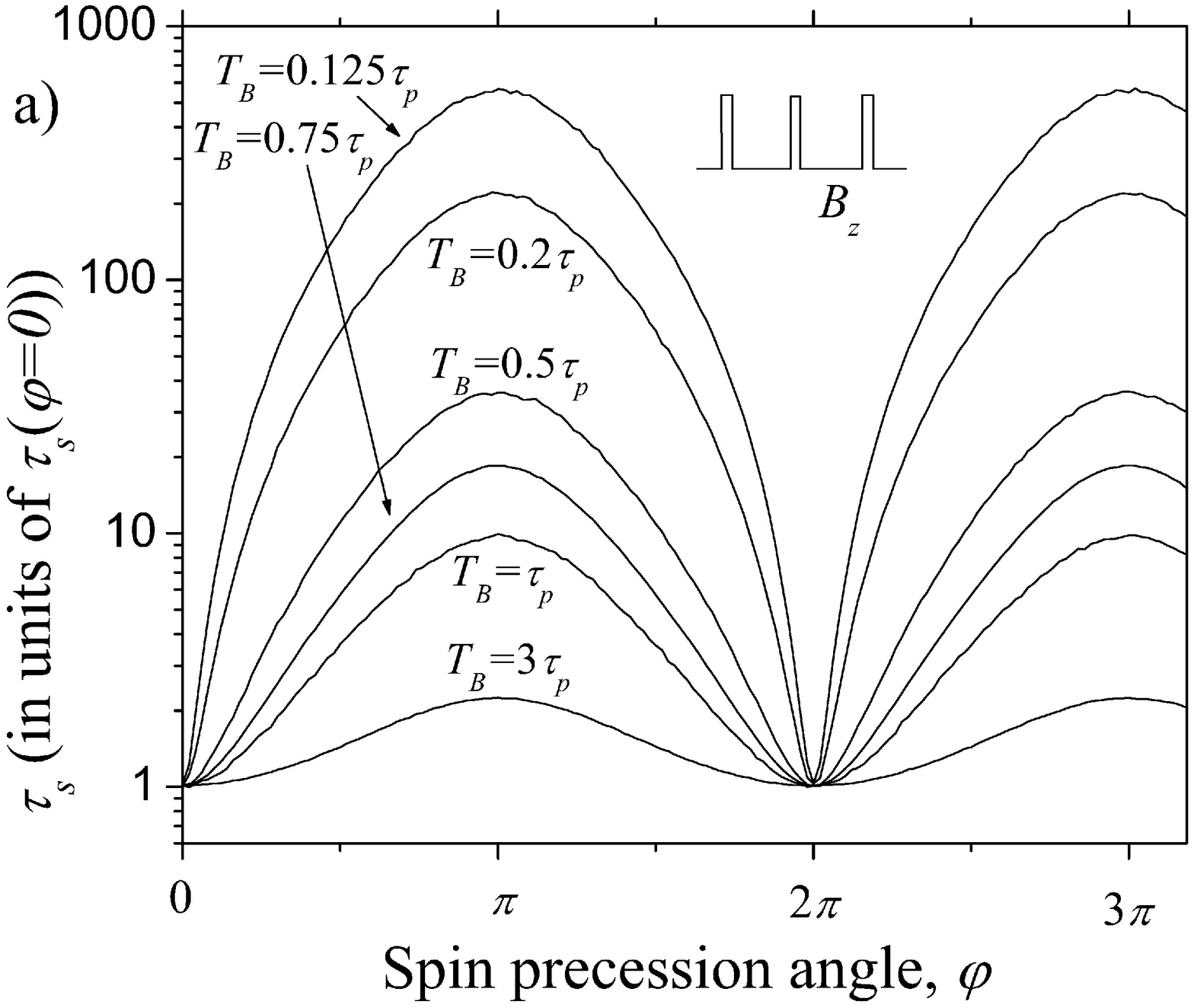}
\includegraphics[height=8cm,angle=270]{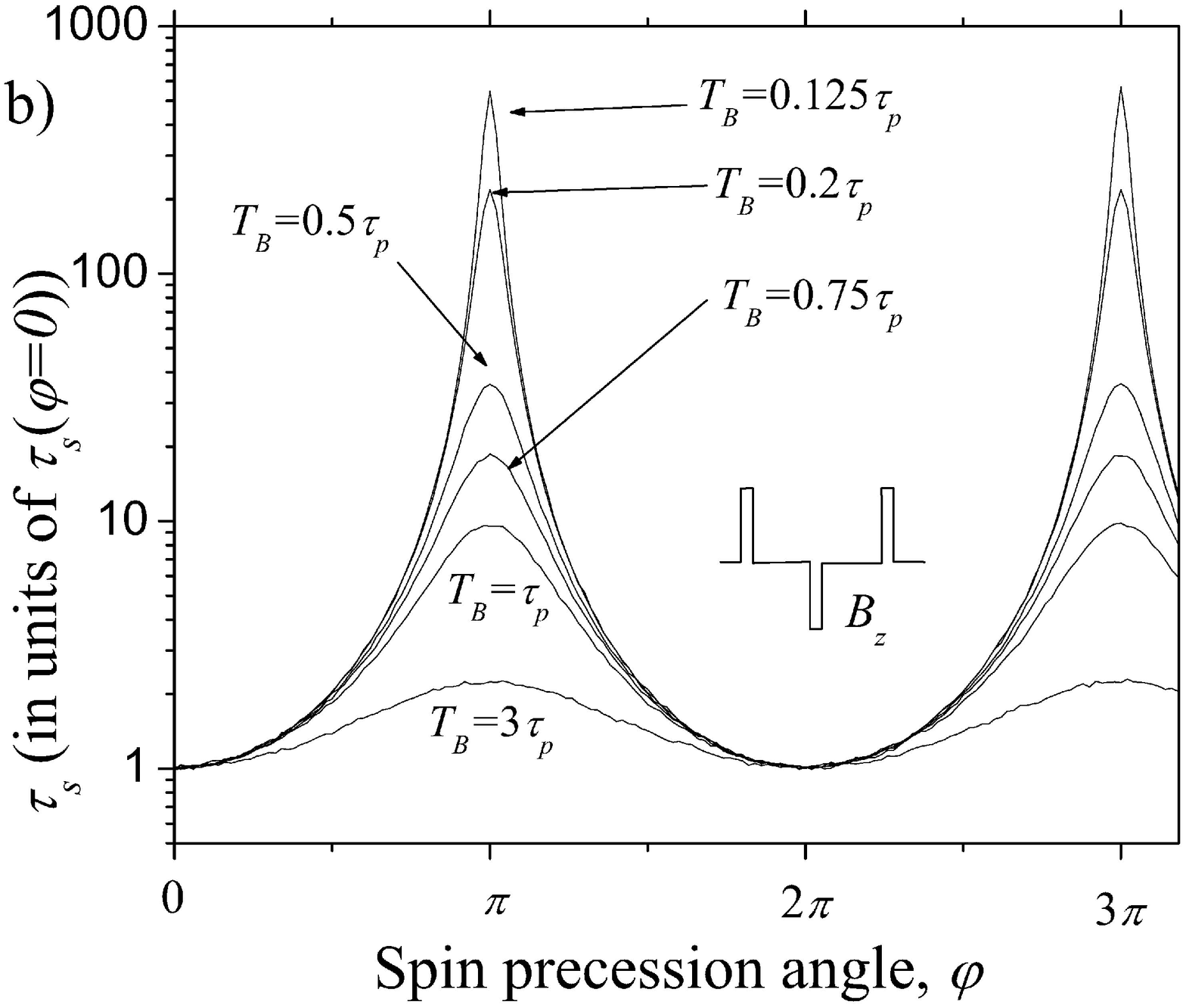}
\caption{Spin relaxation time as a function of the spin precession
angle $\varphi$ due to a pulse for several pulse periods $T_B$ and
two types of pulse sequences: unidirectional pulse sequence (a),
alternating pulse sequence (b). These plots was obtained using the
parameter value $\eta L_p=0.4$.} \label{fig3}
\end{figure}

The time-dependence of $\langle \overrightarrow{S} \rangle$ was
calculated for an ensemble of $10^5$ electrons for each value of
the parameters describing the pulse sequence. The spin relaxation
time for various pulse spacings is shown in Fig. \ref{fig3} as a
function of the spin rotation angle. We found that the rate of
increase of spin relaxation time does not depend on the parameter
$\eta L_p $ when $\eta L_p <1$. Instead, it is completely defined
by the spacing between pulses, by the type of pulse sequence, and
by the spin rotation angle due to a pulse. A strong dependence of
the spin relaxation time on the pulse sequence is observed. For
short spacings between pulses, the unidirectional pulse sequence
suppresses the spin relaxation more efficiently than the
alternating pulse sequence. The spin relaxation time coincides for
both pulse sequences only for $\varphi= \pi n$, where $n$ is an
integer number. Furthermore, the spin relaxation time
$\tau_s(\varphi)$ is a periodic function of $\varphi$ with period
$2 \pi$, symmetric within a period,
$\tau_s(\pi+\beta)=\tau_s(\pi-\beta)$, where $\beta \in [0,\pi]$,
and has a maximum at $\varphi=\pi (2n+1)$. By increasing the
spacing between pulses, the relaxation time decreases for both
sequences. When the spacing between pulses becomes as long as a
few momentum relaxation times, the spin rotations due to
neighboring pulses become uncorrelated and the dependence of the
spin relaxation time on $\varphi$ is the same for both pulse
sequences. This is clearly seen for $T_b=3\tau_p$ in Fig.
\ref{fig3}. It is important to notice that a significant increase
of spin relaxation time is observed even when the spacing between
pulses is longer than $\tau_p$.

\begin{figure}[tb]
\centering
\includegraphics[height=7.5cm,angle=270]{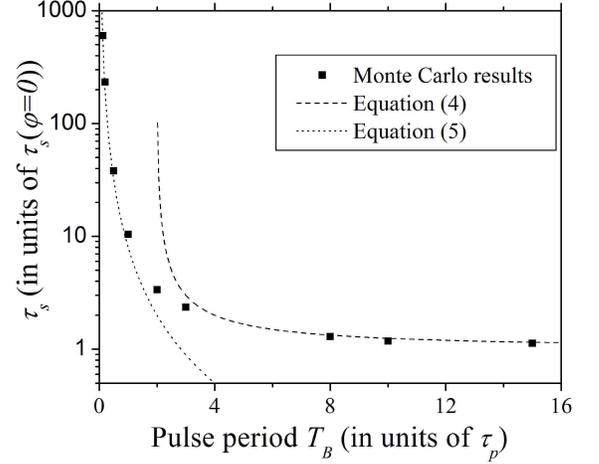}
\caption{Spin relaxation time as a function of the spacing between
pulses $T_B$ at $\varphi=\pi$. The assymptotic behavior of the
spin relaxation time (Eqs. (\ref{eq4}), (\ref{eq5})) is in
excellent agreement with Monte Carlo results.} \label{fig4}
\end{figure}

Fig. \ref{fig4} shows the spin relaxation time as a function of
the spacing between pulses $T_B$ in the practically important
situation $\varphi=\pi$, which is characterized by the longest
spin relaxation time. The spin relaxation time sharply increases
at small values of $T_B$ and slowly decreases with increase of
$T_B$ to the spin relaxation time without pulses
$\tau_s(\varphi=0)$. Let us derive the asymptotic behavior of the
spin relaxation time as a function of spacing between pulses in
this case. First, consider the limit of distant pulses, when $T_B
\gg \tau_p $. Using a method described in Ref. \cite{r35} and
assuming that a pulse is applied in an arbitrary time moment $t$
between two scattering events separated by a time interval $\tau$,
the mean squared dephasing between these scattering events
$\overline{\delta \theta^2}$ is given by

\begin{equation}
\overline{\delta \theta^2}=\frac{1}{\tau}\int\limits_0^\tau
\Omega^2(2t-\tau)^2dt=\frac{1}{3}\Omega^2\tau^2. \label{eq2}
\end{equation}
The mean squared dephasing between two scattering events without a
pulse is simply given by $\overline{\delta
\theta^2}=\Omega^2\tau^2$. Taking into account the pulse
probability, $\tau / T_B$, and the exponential distribution of
probability of scattering,  $p(\tau,
\tau+d\tau)=(1/\tau_p)\textnormal{exp}\left( -\tau /
\tau_p\right)d\tau$, the mean free dephasing after
$\overline{\delta \theta^2}$ scattering events will be

\begin{eqnarray}
\overline{\delta \theta^2}=n\frac{1}{\tau_p}\int\limits_0^\infty
e^{-\frac{\tau}{\tau_p}}\left(\left(1-\frac{\tau}{T_B}
\right)+\frac{\tau}{T_B}\frac{1}{3}\right)\Omega^2\tau^2d\tau=
\nonumber
\\
2n\Omega^2\tau^2_p\left(1-2\frac{\tau_p}{T_B} \right). \label{eq3}
\end{eqnarray}
If we take the relaxation time $\tau_s$ for a group
of spins in phase at the initial moment of time to get about one
radian out of step, we find
\begin{equation}
\tau_s=\frac{\tau_p}{2\Omega^2 \tau_p^2}
\frac{1}{1-2\frac{\tau_p}{T_B}} \label{eq4}.
\end{equation}

The first term at the right-hand side of Eq. (\ref{eq4}) is the
spin relaxation time without pulses, the second term describes the
effect of the pulse sequence. In the opposite limit, when the
number of pulses per mean free path is large $T_B \ll \tau_p$, the
spacing between pulses $T_B$ defines the characteristic angle of
spin precession between two scattering events, instead of the
momentum relaxation time $\tau_p$. Thus we can write
\begin{equation}
\tau_s \sim \frac{\tau_p}{\Omega^2 T_B^2} \label{eq5}.
\end{equation}
The asymptotic expressions for spin relaxation time, Eqs.
(\ref{eq4}), (\ref{eq5}) are presented in Fig. \ref{fig4} showing
an excellent agreement with Monte Carlo results.

We would like to emphasize that the proposed technique is most
suitable for clean quantum structures with low electron density at
low temperatures, i.e. when $\tau_p$ is long. For example, taking
$v_F=5 \cdot 10^6$cm/sec and $L_p=1\mu$m we obtain $\tau_p=20$ps.
Our calculations indicate that in order to get a two-fold increase
in $\tau_s$, the spacing between the pulses at $\tau_p=20$ps
should be $\sim 50$ps at $\varphi =\pi$. The calculations
presented in this paper have been made for a particular value of
the parameter $\eta L_p=0.4$. This specific value of $\eta L_p$ is
realizable in physical systems. For instance, considering  an
InAlAs/InGaAs quantum well \cite{Nitta} with $\alpha=0.4\cdot
10^{-12}$ eV m, $m^*=0.04m_e$, and $L_p=1 \mu$m, we obtain $\eta
L_p=0.42$. We would like to emphasize again that the {\it rate} of
change of $\tau_s$ does not depend on $\eta L_p$ in the
motional-narrowing regime.

In order to experimentally observe supression of spin relaxation,
the energy of below-band gap laser must be adjusted to minimize
the excitation of real carriers by compromising between
state-filling effects and magnitude of the Stark shift
\cite{awschalom}. Recent calculations for quantum dot geometry
demonstrate that $\pi$ pulses may be obtained even for quite large
detunings ($\sim 70$meV) and experimentally realistic pump
parameters. Definitely, the role of unwanted carriers excitation
is smaller in systems with $\eta L_p \gtrsim 1$, when just several
pulses significantly increase $\tau_s$. Another important effect
that should be avoided is the sample heating. This can be done in
the following way. The initial electron spin polarization can be
excited $N$ times per second and followed by $M$ below-band gap
pulses, so that the total number of pulses per second is $N \cdot
M$. Typically, $1$ns$\lesssim\tau_s(\varphi=0)\lesssim 100$ns.
Therefore, the required number of pulses $M$ to monitor
enhancement of $\tau_s(\varphi=0)$ at $T_B=50$ps is $\sim
\tau_s(\varphi=0)/T_B=20..20000$.
 In the recent experiment \cite{awschalom} the
sample was not significantly heated at $250$kHz pump repetition
rate. Consequently, the sample heating is smaller than in Ref.
\cite{awschalom} if $N< [250000/M]$, where $[..]$ denotes the
integer part.

In conclusion, the electron spin relaxation due to the
D'yakonov-Perel' spin relaxation mechanism under a sequence of
optical below-band gap pulses was studied. The pulse sequence
rotating electron spins around the axis perpendicular to the
quantum well significantly suppress spin relaxation in a way quite
similar to the spin echo in the nuclear magnetic resonance. The
spin rotation mechanism is based on the optical Stark effect. It
was demonstrated that the optical Start effect in semiconductors
allows obtaining very short ($\sim 150$fs) and strong ($\sim 20
$T) pulses of effective magnetic field \cite{awschalom}. Spin
relaxation time was calculated for different pulse sequences and
spacing between pulses. It was found that, in general,
unidirectional pulse sequences suppress the spin relaxation more
efficiently than the alternating one. Analytical formulae for
asymptotic behavior of the spin relaxation time were obtained. The
proposed method of spin coherence control could find applications
in probing spin-coherence dynamics in heterostructures.

We gratefully acknowledge helpful discussions with C.
Piermarocchi, V. Privman and S. Saikin. This research has been
initiated at the Center for Quantum Device Technology at Clarkson
University and was supported by the National Science Foundation,
grant DMR-0312491.

\end{document}